\documentclass[ reprint,
                superscriptaddress,
                groupedaddress,
                amsmath,amssymb,
                aps,
                prl,
                citeautoscript,
                showkeys
              ]{revtex4-1}

\usepackage[utf8]{inputenc}

\usepackage{amsmath}
\usepackage{amsfonts}

\usepackage[caption=false]{subfig} 
\usepackage{graphicx}

\usepackage{pgfplots}
\pgfplotsset{compat=1.14}
\usetikzlibrary{arrows}
\usetikzlibrary{backgrounds}
\usetikzlibrary{calc}
\usetikzlibrary{decorations.text}
\usetikzlibrary{decorations.markings}
\usetikzlibrary{external}
\usetikzlibrary{fit}
\usetikzlibrary{intersections}
\usetikzlibrary{mindmap}
\usetikzlibrary{positioning}
\usetikzlibrary{shapes.geometric}
\usetikzlibrary{shapes}
\usepackage{xspace}
\usepackage{standalone}
\usepgfplotslibrary{groupplots}
\usepgfplotslibrary{fillbetween}

\usepackage{siunitx}
\usepackage{hyperref}

\usepackage[capitalise]{cleveref}

\usepackage{todonotes}
\usepackage{soul}

\newcommand{\cf}{cf.\xspace}
\newcommand{\eg}{e.\,g.\xspace}

\newcommand{\thomsonlab}
{
    Division of Biology and Biological Engineering,
    California Institute of Technology,
    Pasadena, California 91125,
    USA
}

\newcommand{\breakuptime}{\ensuremath{\tau_b}\xspace}
\newcommand{\activationarea}{\ensuremath{\mathcal{A}}\xspace}
\newcommand{\radius}{\ensuremath{r}\xspace}
\newcommand{\Cutoff}{\ensuremath{R_c}\xspace}
\newcommand{\cutoff}{\ensuremath{r_c}\xspace}
\newcommand{\windowsizedef}{\ensuremath{t_{w}=0.5}\xspace}

\newcommand{\vcirc}{\ensuremath{u}\xspace}

\begin{document}

%
%


\title{Phenomenological model of motility by spatiotemporal modulation of active interactions}

\author{Dominik Schildknecht}
\email{dominik.schildknecht@gmail.com}
\affiliation{\thomsonlab}
\author{Matt Thomson}%
\email{mthomson@caltech.edu}
\affiliation{\thomsonlab}

\date{\today}



\begin{abstract}
    Transport at microscopic length scales is essential in biological systems and various technologies, including microfluidics. Recent experiments achieved self-organized transport phenomena in microtubule active matter using light to modulate motor-protein activity in time and space. Here, we introduce a novel phenomenological model to explain such experiments. Our model, based on spatially modulated particle interactions, reveals a possible mechanism for emergent transport phenomena in light-controlled active matter, including motility and contraction. In particular, the model's analytic treatment elucidates the conservation of the center of mass of activated particles as a fundamental mechanism of material transport and demonstrates the necessity of memory for sustained motility. Furthermore, we generalize the model to explain other phenomena, like microtubule aster-aster interactions induced by more complicated activation geometries. Our results demonstrate that the model provides a possible foundation for the phenomenological understanding of light-controlled active matter, and it will enable the design and optimization of transport protocols for active matter devices.
\end{abstract}
\maketitle
\section{Introduction} \label{sec:intro}
Transport at micron length scales is essential in technologies such as microfluidics~\cite{Kuo2011,Convery2019} and many biological processes, including cell motility~\cite{Mitchison1996,Blanchoin2014}, cell division~\cite{Fletcher2010}, and intracellular material transport~\cite{Hirokawa2009,Appert-Rolland2015}.
While in most technological applications, transport arises as the response of a system to externally applied forces, biological systems rely on molecular self-organization to achieve material transport. Because the biological systems are intrinsically out of equilibrium, transport can not be explained by linear response theory~\cite{DalCengio2019}. This property makes these systems fascinating to study and much more challenging to understand. Indeed, there have been various \textit{in vivo} and \textit{in vitro} studies trying to address biological transport phenomena at the micron length scales. However, elucidating biological transport mechanism remains challenging due to the non-linear, self-organized nature of transport and the ``molecular complexity'' in biological systems~\cite{Mitchison1996,Hirokawa2009,Fletcher2010,Blanchoin2014,Appert-Rolland2015}.

In order to understand the physics of biological systems in more detail, so-called active matter was introduced. Active matter systems continuously transform energy to power various intrinsic processes such as self-propulsion in particle-based active matter or motor-protein movement and polymerization in protein-based active matter (see, \eg, Ref.~\cite{Needleman2017} for a review). One area of active matter research focuses on the behavior of different classes of active Brownian particles~\cite{Howse2007,Hong2007,Jiang2010a,Romanczuk2012,Cates2013}. These systems have been examined, for example, to understand phenomena such as the emergent phase separation arising from repulsive interactions~\cite{Theurkauff2012,Buttinoni2013,Cates2015}. More recently, spatiotemporal modulation of the single-particle velocity (representing the activity in these systems) was introduced~\cite{Walter2007,Volpe2011,Buttinoni2012,Fragkopoulos2020}, enabling the construction of microfluidic devices through self-assembly~\cite{Stenhammar2016}.

In general, particle-based active matter systems focus on the emergent dynamics of individual agents, modeling the dynamics of interacting populations of molecules, cells, or organisms. In this area, active Brownian particles successfully describe the emergent phenomena, even though their interactions are predominantly repulsive~\cite{Cates2015}. However, the predominantly repulsive and short-range nature of particle interactions in these models limits the self-organization capability making it challenging to assemble structures with long-range order, like the mitotic spindle, that occur in cells. In contrast, protein-based active matter systems, realized experimentally by mixtures of microtubules and motor proteins~\cite{Urrutia1991,Nedelec1997,Surrey2001}, spontaneously organize structures including contractile networks, microtubule asters, and nematic phases that exhibit long-range order. Using these protein-based active matter systems, one can systematically analyze how self-organization can lead to structures such as vortices and asters. Furthermore, the complex organizational behavior observed in these systems inspired interesting modeling approaches that rely on a hybrid between hydrodynamics and active nematicity~\cite{Marenduzzo2007,Marenduzzo2007a,Giomi2008,Giomi2015}. Self-organization phenomena occurred both in experiments and models~\cite{Nedelec1997,Surrey2001,Giomi2012,Giomi2014}, but because the chemical reactions happen everywhere, they are challenging to isolate and inspect in detail. To focus on the relevant mechanisms, localization of the chemical reactions is imperative. The required localization was recently achieved by introducing light-activated reversible linking between the kinesin motors, leading to a spatially confined contraction of the microtubule network~\cite{Ross2019}. 

The excellent spatial control of the contractile microtubule network was then used to understand how motility can arise in biological systems. In particular, Ross \textit{et al.}~\cite{Ross2019} demonstrated that the directed motion of asters could emerge through the spatiotemporal variation of kinesin motor activation. Specifically, the size of the light-activated region determines the size of the aster, and the activation pattern's trajectory through space controls the asters' trajectory. Furthermore, the authors of Ref.~\cite{Ross2019} analyzed systems beyond isolated asters to understand the aster-aster interactions. They achieved this interaction by linking two aster-forming regions by a thin connection, leading to the merger of the two asters in a region where asters would otherwise not form. Therefore, these experiments provided many insights into the physics of material transport and contractile networks and laid the foundation for the applications of such systems. However, the underlying physical phenomena are not yet well understood, so that a theoretical model is required to deepen our understanding of motility in biological systems.

This paper aims to address this knowledge gap by introducing a phenomenological model of material transport in spatiotemporal-activated contractile networks by a particle-based description. Our model is based on spatiotemporal control over interactions between the particles. In particular, we will consider attractive harmonic interactions motivated by continuum descriptions~\cite{Foster2015a,Qu2021b}.
We then use a simple limit of the model to show that the fundamental mechanism of transport can be understood in terms of the conservation of the center of mass of activated particles. Furthermore, we demonstrate that in our model, memory in the form of retained activation is required to describe the experimentally observed long-term stabilization of asters~\cite{Ross2019}. In particular, we will express the linear dependence between the speed of the activation region and the trailing distance of the aster in microscopic parameters. In addition, aster-aster interactions are simulated using the linking of two aster-forming regions by an activation ``bridge,'' leading to qualitatively similar results between our phenomenological model and the experimental results. In particular, we also observe the exponential decay of the aster-aster distance with time, analogous to the experimental results~\cite{Ross2019}. 

\section{Model} \label{sec:model}
\begin{figure}
    \centering
        \includegraphics[width=\columnwidth]{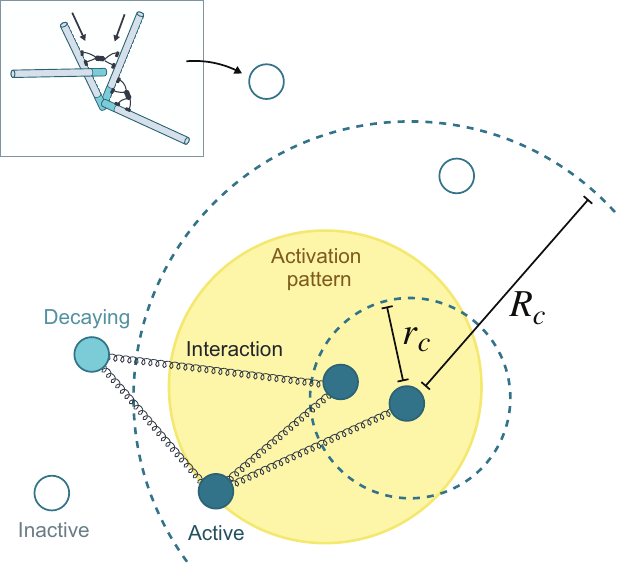}
        \caption{Sketch for the model outlined in this paper: Particles can be thought of as ``mini-asters'' already consisting of a large number of microtubules and motors, averaging to an isotropic point particle. Particles can be activated (blue or light blue) or inactivated (white). Particles outside of the activation region (yellow) lose their activation state in a characteristic time of $\lambda^{-1}$ (hence displayed in light blue). All activated particles interact with each other through a spring potential if their pairwise distance is larger than $\cutoff$ and smaller than $\Cutoff$. Intuitively, the interaction between particles occurs because microtubules of mini-asters overlap, and when the kinesin is activated to cross-link, these overlapping microtubules pull on each other in a telescoping-like movement to form an attractive interaction~\cite{Foster2015a}.}
    \label{fig:model-sketch}
\end{figure}

Here, we propose a coarse-grained model of the experimental system that collects small clusters of microtubules and motors into $N$ point-like particles in two dimensions, which neglects the sub-structure of each cluster~\cite{Foster2015a}. For simplicity, a sketch of the model is provided in \cref{fig:model-sketch}. In this model, a particle $j$ is described by its position $\vec x_j$, velocity $\vec v_j$, and activation state $p_j$, which is either $1$ or $0$ for activated or inactivated particles, respectively. The interactions arise in our model, since microtubules of one particle overlap with the microtubules of neighboring particles. If the two particles are in an activated state, then the kinesin motor is expected to bind the microtubules together leading to an effective attraction by the kinesin moving in opposite directions on the (pre-polarized) microtubules of different micro-asters. Therefore, only activated particles interact and their effective one time-step equation of motion is
\begin{subequations}
    \begin{align}
        \vec x_i(t+\Delta t) &=  \vec x_i (t)+\Delta t\, \vec v_i(t),  \label{eq:eom:x}\\
        \vec v_i(t+\Delta t) &=  \frac{\Delta t}{m_i} \vec F^{(a)}_i\left[\{\vec x_j\}_j\right],  \label{eq:eom:v}\\
        \vec F^{(a)}_i\left[\{\vec x_j\}_j\right] &= -\nabla_{i} \left[ \frac{k}{2}\sum_{\substack{i\neq j\\|\vec x_i-\vec x_j|>\cutoff\\|\vec x_i-\vec x_j|<\Cutoff}}p_i p_j(\vec x_i-\vec x_j)^2 \right].  \label{eq:eom:fp}
    \end{align}
    \label{eq:eom}%
\end{subequations}
These equations arise as an explicit Euler update for the Newtonian equations of motion. These equations can be contrasted to the typical description of drag-dominated systems, which neglect the inertial term~\cite{Chaikin1995Ch7,Howard2001Chapter2}. Indeed, we will observe in our results that the effective mass of emergent structures can grow so that the inertial term should not be neglected. Hence, the term is kept, and the update rule \cref{eq:eom:v} is derived from an explicit Euler update for the velocity of a particle with drag $\gamma$ under the influence of a force $\vec F$: $\vec v_i(t+\Delta t)=(1-\gamma\Delta t/m_i)\vec v_i(t)+\Delta t/m_i \vec F$. In this standard update, the first term requires a time step so that $1\gg \gamma\Delta t/m_i $, since otherwise, the explicit Euler scheme is divergent. For large drag, this requires very small time steps. However, the first term's effect is merely to diminish the velocity exponentially fast. Therefore, we can use the separation of time scales to remove the first term in the large drag limit, where the velocity decays instantaneously. While \cref{eq:eom} arises from Newton's equations in conjunction with time-scale separation, the equations are indeed equivalent to neglecting the inertial term~\cite{Chaikin1995Ch7,Howard2001Chapter2}. However, in this interpretation, the factor $\Delta t/m_i$ in \cref{eq:eom:v} should be replaced by $\gamma_i^{-1}$.

In \cref{eq:eom:fp}, it is evident that the only force in our model is an active force $\vec F^{(a)}$, but one could also incorporate additional thermal forces $\vec F^{(th)}$ here. In particular, $\vec F^{(th)}$ could be modeled by a Gaussian random process with zero mean and correlations $\langle F^{(th)}_\mu(t)F^{(th)}_\nu(t')\rangle\propto T\delta_{\mu\nu}\delta(t-t') $ where the subscripts denote vector components, and $T$ is the temperature~\cite{Chaikin1995Ch7}. However, as demonstrated in Ref.~\cite{Ross2019} (in particular Fig.~1c and SI 2.10), diffusion does not affect the formation of asters, but only their decay once no more light activation is provided. Hence, for this paper, where we want to describe the emergent motility in microtubule aster systems, no additional thermal forces will be used. 

Therefore, we only incorporate one force, namely the active force $\vec F^{(a)}$, which depends on the set of all position vectors $\{\vec x_j\}_j$. The force is derived from a spring potential with spring constant $k$ and vanishing rest length. Using the spring potential to describe the active interactions is in agreement with both the experiment~\cite{Ross2019} and continuum-based theoretical descriptions~\cite{Foster2015a,Qu2021b}. The interaction only affects particle pairs with pairwise distances between \cutoff and \Cutoff. The upper truncation \Cutoff ensures that arbitrarily distant particles do not interact. For this paper, where we only treat localized activated particles, the upper cutoff is mostly irrelevant, so that we use $\Cutoff\rightarrow \infty$. However, if larger systems with more complicated activation patterns are considered, the upper truncation \Cutoff has to be chosen more carefully. Additionally, a small lower truncation \cutoff is enforced in the simulations to avoid numerical instabilities. 

Finally, spatiotemporal control of activity is introduced in the following way: A particle $j$ becomes activated at time $t$ if $\vec x_j\in\activationarea (t)$, where $\activationarea (t)$ describes the activation area due to the external stimulus, such as the light-activation pattern in the experimental system~\cite{Ross2019}. In the most elementary version of the model, a particle will deactivate immediately after leaving \activationarea. However, later in this paper, the influence of retained activity will be analyzed. There, the deactivation is not immediate but is instead governed by a Poisson point process, leading to an exponential decay of the probability of being activated with activation decay rate $\lambda$. The elementary model is retrieved for $\lambda\rightarrow \infty$ so that each particle's memory becomes infinitely short, and they indeed deactivate immediately after leaving \activationarea. 

In this paper, the discussion is restricted to two experimentally analyzed situations~\cite{Ross2019}: First, \activationarea consists of a single circle with radius $\radius$ moving in a straight line with velocity \vcirc. Second, \activationarea consists of two stationary circles with radius $\radius$ separated by a distance of $2l$, connected by a small connection of width $w$.

\section{Analysis of the emergent aster motility mechanisms} \label{sec:ma}
First, we focus on the moving circle protocol, which has experimentally led to a motile aster following the activation circle~\cite{Ross2019}. For the numerical integration of the model, the following (non-dimensional) values are used if not specified differently: The system consists of $4960$ particles in a square-shaped box with an area of $4^2$. The box is co-moving with the activation area with radius $\radius=0.5$, and particles are added and removed from the simulation to keep a constant density of $310$ particles per unit area. In particular, the particles are added in the (non-activated) boundary region that appears new to field of view of the simulation, and removed from the area that falls out of the field of view of the simulation. In addition, activated particles that are close are fused, in order to ensure a constant number of particles, eventhough the mass of all particles in the field of view is increased. This system is integrated for $1000$ time steps with $\Delta t = 0.01$. For stability, a lower truncation of $\cutoff=5\cdot 10^{-3}$ is used. The mass of each particle and the spring constant are $m_i=k=1$ for simplicity [leading to $\gamma_i=100$ in the formulation of \cref{eq:eom:v} in terms of drag]. Initially, the situation without memory ($\lambda\rightarrow \infty$) will be discussed.

\subsection{Aster motility and breakup for memoryless particles} \label{sec:ma:no-memory}
\begin{figure}
    \centering
        \includegraphics[width=\columnwidth]{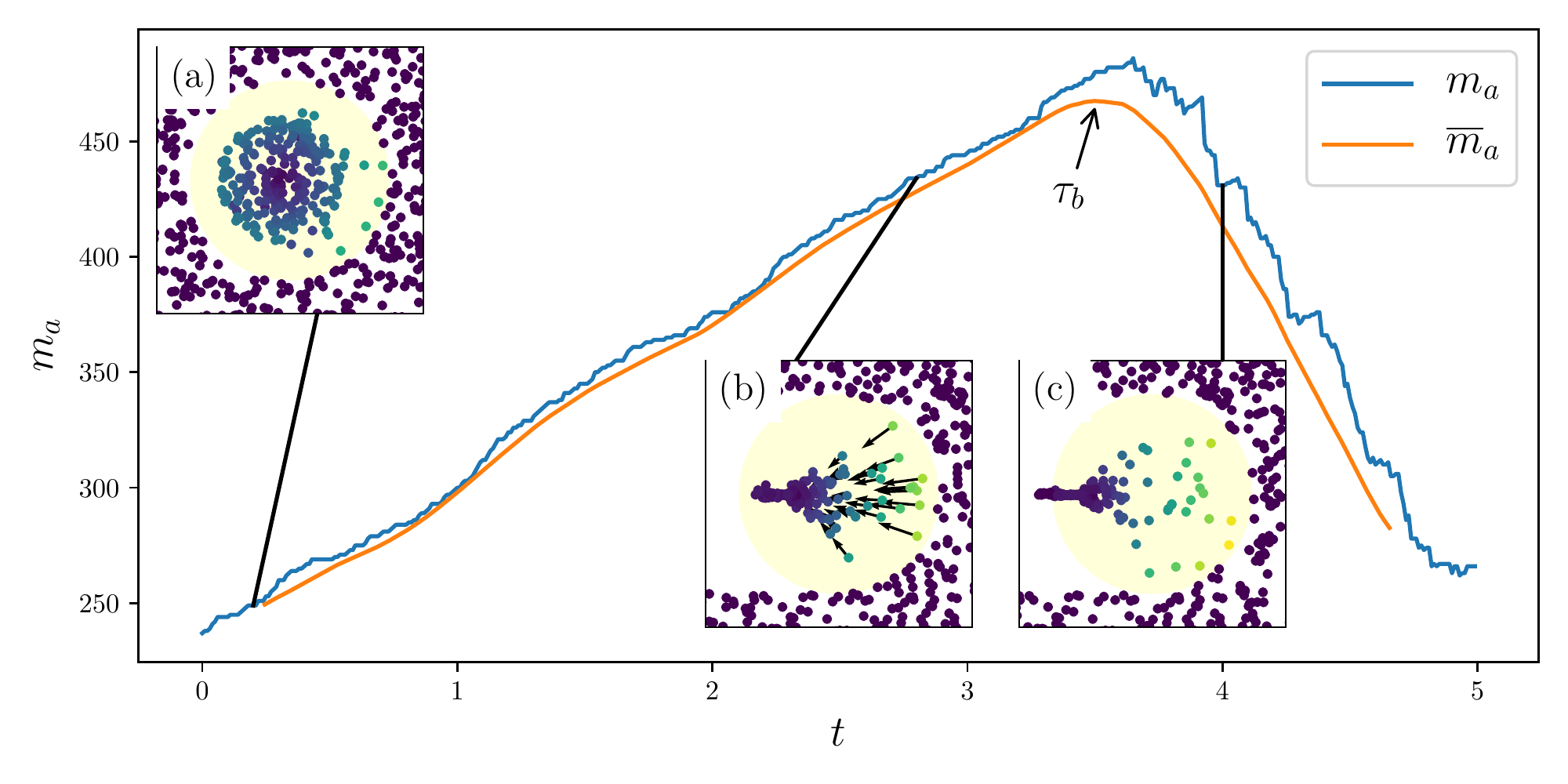}
        \caption{Time evolution of the activated mass $m_a(t)$ and its moving average $\overline m_a(t)$ for a system with $\vcirc=0.2$. The curve is annotated with illustrations of the configurations: The particles are color-coded according to their velocity, where violet indicates standing still, blue moving slow, and yellow moving fast. The activation region is highlighted in faint yellow. Furthermore, inset (b) additionally depicts the particle velocities, illustrating the merging process. It should be noted that the depletion of particles to the left of the activation area in (b) and (c) is due to the history of the system: In these areas, the activation already occurred and particles are absorbed in the aster. Hence the trace of the activation area can be seen in the data as depletion areas.}
    \label{fig:te}
\end{figure}

The typical time evolution of such a system is presented in \cref{fig:te} for the first $500$ time steps. Specifically, the time dependence of the mass of activated particles $m_a$ is shown, annotated with illustrations of the configurations. At the beginning of the simulation, the mass inside the initial circle $\activationarea (t=0)$ is activated, which contracts to form a massive core, as observed in the inset (a) of \cref{fig:te}. In analogy with experiments~\cite{Ross2019}, the massive core will be referred to as an aster. Subsequently, material in front of the aster is newly activated, fusing with the aster leading to an increase in $m_a$ and an effective mass transport by shifting the center of mass [\cf inset (b) in \cref{fig:te}]. This process is sustained until the aster becomes too massive to be moved by the newly activated material. Finally, the aster leaves the activation area and breaks apart, as seen in the inset (c) in \cref{fig:te}, which leads to a rapid drop in activated mass.

We will observe that such breakup events occur for various values of system parameters. However, the time at which they occur can vary strongly. To quantify the breakup behavior, the time until the first significant drop in activated mass is considered. Intuitively, the longer it takes until the rapid drop occurs, the more stable is the mass transport. Here, the breakup time \breakuptime is defined as the first time that the activated mass $m_a$ decreases significantly. Formally,
\begin{align}
    \breakuptime = \min_{t>0} \left\{ \; t\; \middle|\frac{ \mathrm d }{\mathrm d t}\,\overline m_a (t)<0 \right\},
    \label{eq:def-breakuptime}
\end{align}
where $\overline m_a$ is the (centered) moving average of $m_a$. The moving average is required due to fluctuations arising in simulations. A window size for the moving average of \windowsizedef was used, but variations of the window size only affected the results marginally.

\begin{figure}
    \centering
    \subfloat[][]{
        \includegraphics[width=.8\columnwidth]{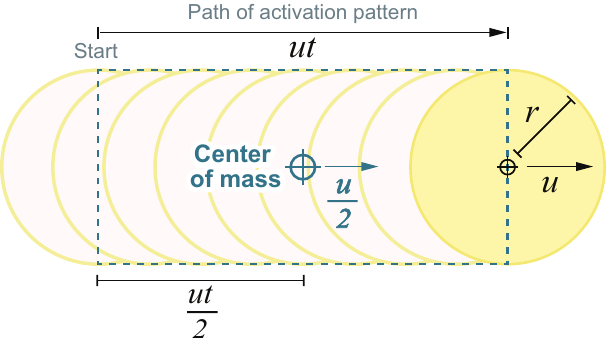}
        \label{fig:sketch}
    }\\
    \subfloat[][]{
        \includegraphics[width=\columnwidth]{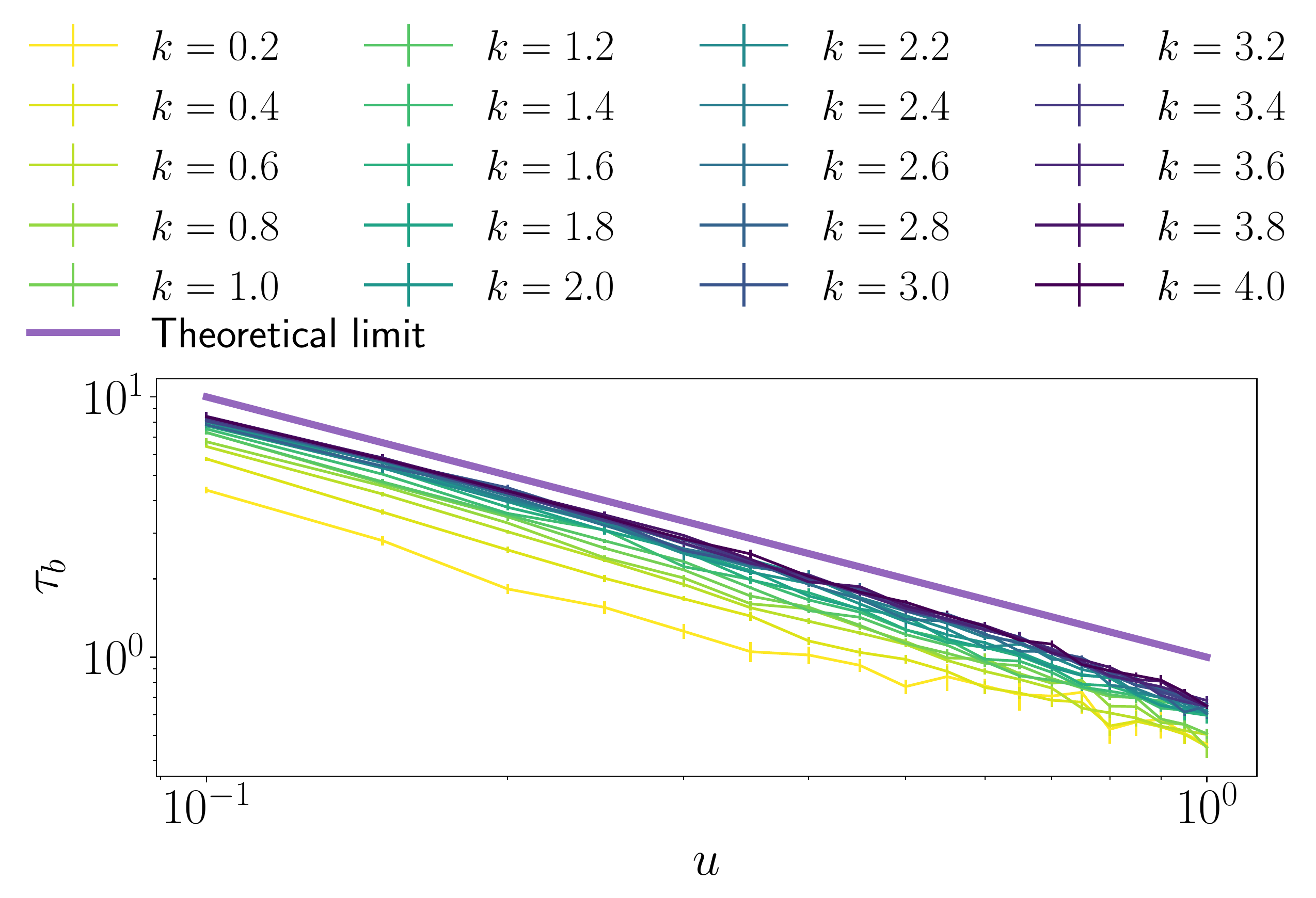}
        \label{fig:spring-k}
    }
    \caption{In (a), the analytic calculation is exemplified: The faint circles indicate the path of the activation pattern, and the yellow circle is the activation area at time $t$. Due to symmetry, the aster travels at half the speed of the activation region. In (b), the breakup time \breakuptime from simulations as a function of activation-pattern velocity \vcirc for various spring constants $k$ is compared to the theoretical limit. Inverse proportionality can be observed for all $k$, where the curves for larger $k$ asymptotically approach the theoretical limit given in \cref{eq:breakuptime-infinite-k}.}
\end{figure}

For an intuitive understanding of the fundamental mechanisms of aster motility and the breakup time, the sketch in \cref{fig:sketch} can be considered, which illustrates the moving activation pattern. In the limit of very large spring constants, the timescale for contraction is much faster than the timescale associated with moving the activation pattern. In this limit, the collision of the aster and the newly activated mass sickle is instantaneous so that no activated particle is lost and all activated mass is accumulated at a single point. Because the system's momentum vanishes, the center of mass remains invariant, and all mass is collected at the center of (activated) mass. For a constant density of particles, the center of mass is at the center of the area that was activated at any previous time. According to the geometry presented in \cref{fig:sketch}, the center of mass's velocity is, therefore, half of the activation area's velocity. Without memory, the breakup occurs immediately once the center of mass leaves the activation area \activationarea, therefore,
\begin{align}
    \breakuptime = 2\frac{\radius}{\vcirc}
    \label{eq:breakuptime-infinite-k}
\end{align}
so that the aster travels exactly $d=2\radius$ before breaking apart~\footnote{in \cref{fig:sketch}, the aster would have already broken apart}. For smaller spring constants, the aster will lag behind the center of mass, leading to an earlier breakup. Therefore, the expression in \cref{eq:breakuptime-infinite-k} is an upper limit for the breakup time of memoryless particles.

This prediction is indeed validated by the simulation results for finite spring constants presented in \cref{fig:spring-k}. Here, the curves converge to the theoretical limit for increasing spring constants. Furthermore, all \breakuptime curves exhibit the inverse proportionality with respect to the velocity, generalizing the prediction of \cref{eq:breakuptime-infinite-k} to finite $k$.

\subsection{Sustained aster motility by inclusion of memory}\label{sec:ma:with-memory}
While the simulation results presented in \cref{fig:spring-k} can be well understood in the theoretical framework, it is not representative of the experiments~\cite{Ross2019}. Specifically, the aster remains stable even after it leaves the activation area. This discrepancy between our simple calculation and the experiment can be explained as in the experiments the disassociation of motors is not instantaneous upon exiting the optical activation area. In our model, we can introduce such retained activation by deactivating particles in a Poisson point process, which surmounts in a finite (rather than infinite) activation decay rate $\lambda$. Therefore, the particles do not deactivate immediately upon leaving the activation area \activationarea, but their probability of being activated decreases exponentially with a time constant $\lambda^{-1}$. 

Before this extended model is simulated, the effect of activation retention will be demonstrated in the limit of infinitely strong interactions, similar to the calculation leading to the breakup time \breakuptime in \cref{eq:breakuptime-infinite-k}. In the initial phase, the aster contracts and stays inside the activation area \activationarea. In this phase, the behavior of the system is independent of the value of $\lambda$ since the particles cannot deactivate while in \activationarea. Because of the infinitely strong interactions, all mass is collected by the aster. Therefore, the aster's mass increases linearly with a rate of $2r\vcirc \rho$, where $2r\vcirc$ is the newly activated area per unit time, and $\rho$ is the area mass density, i.e., the mass per unit area. After the initial phase, where the aster is in the activation area \activationarea, activated particles continuously deactivate with rate $\lambda$. Then, assuming that the aster does not reenter the activation area, the mass is described by the differential equation
\begin{subequations}
    \begin{align}
        \frac{\mathrm d}{\mathrm d t} m(t)&= -\lambda m(t)+2r\vcirc\rho,\\
        m(t=\breakuptime)&= (\pi+4)\rho r^2,
        \label{eq:ode-mass-phase2:initial}%
    \end{align}
    \label{eq:ode-mass-phase2}%
\end{subequations}
where the initial condition \cref{eq:ode-mass-phase2:initial} arises due to the mass in the initial circle and the linear growth that happens until \breakuptime. \Cref{eq:ode-mass-phase2} has the solution
\begin{align}
    m(t\geq\breakuptime) &=  \frac{2r\vcirc\rho}{\lambda} + \frac{r \rho }{\lambda}\left[ (4+\pi)r\lambda-2\vcirc \right]e^{-\lambda (t-\breakuptime)},
    \label{eq:sol-mass-phase2}
\end{align}
describing an exponential convergence to the asymptotic limit $m(t\rightarrow \infty)=2r\vcirc\rho\lambda^{-1}$ after the aster leaves the activation area. Since $m(t)>0$ for all times $t>0$, the aster will not decay completely, but it will remain to have a finite mass. Therefore, we have shown that activation retention can lead to a (quasi-)stable aster, such as those observed in the experiments~\cite{Ross2019}.

Furthermore, also the experimental observation of linear scaling of the distance between the aster and the activation area~\cite{Ross2019} (the so-called trailing distance) can be understood: One can consider the aster's position $x(t)$ for times $t>\breakuptime$, which can be computed by considering a small time window $\Delta t$, where the aster fuses with the newly activated sickle at the common center of mass. Then, the aster's position at time $t+\Delta t$ is
\begin{align}
    x(t+\Delta t)&= \frac{m(t) x(t) + \Delta m\, x^{\text{Sickle}}(t)}{m(t)+\Delta m}, 
    \label{eq:aster-position-start}
\end{align}
where $\Delta m= 2r\vcirc\rho \Delta t$ is the mass of the sickle, $x^{\text{Sickle}}(t)=(t+\frac{\Delta t}{2})\vcirc+\frac{2}{\pi}r$ is the center of mass of the sickle, and $m(t)$ is the mass of the (growing) aster. Therefore, the aster's position fulfills the differential equation
\begin{align}
    \frac{\mathrm d x}{\mathrm dt} = \lim_{\Delta t\rightarrow 0}\left[ \frac{x(t+\Delta t)-x(t)}{\Delta t} \right]
    &= \frac{2 r \vcirc\rho}{m(t)}\left[ \vcirc t+\frac{2r}{\pi}-x \right],
    \label{eq:aster-position-ode}
\end{align}
with the boundary condition $x(\breakuptime)=2r$. \Cref{eq:aster-position-ode} can be solved analytically under the assumption that the aster does not reenter the activation area so that the solution in \cref{eq:sol-mass-phase2} for $m(t)$ is valid at all times $t\geq\breakuptime$. While the complete solution to \cref{eq:aster-position-ode} is not very insightful, the long-time limit of trailing distance, i.e., the distance the aster is behind the center of the activation circle, has the simple form
\begin{align}
    \lim_{t\rightarrow \infty}\left[ \vcirc t-x(t)\right]&= \frac{\vcirc }{\lambda} - \frac{2r}{\pi}, 
    \label{eq:trailing-distance}
\end{align}
which is in excellent agreement with experimental results presented in Fig.~2c of Ref.~\cite{Ross2019}. In particular, we also observe a linear dependence between the trailing distance and the activation region speed. Even more so, our model predicts that the proportionality constant is $\lambda^{-1}$. Hence, the linear fit in Fig.~2c of Ref.~\cite{Ross2019} can be used to estimate $\lambda^{-1}\approx \SI{2}{\minute}$, which is slightly larger than the value expected in Ref.~\cite{Ross2019} ($\lambda^{-1}_{\text{Ref.~\cite{Ross2019}}}\approx\SI{30}{\second}$), but still comparable in size.

\begin{figure}[!htbp]
    \centering
    \subfloat[][]{
        \includegraphics[width=.7\columnwidth]{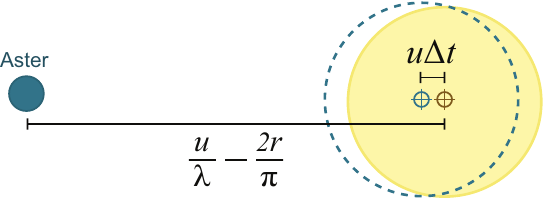}
        \label{fig:stabilization:sketch}
    }\\
    \subfloat[][]{
        \includegraphics[width=.8\columnwidth]{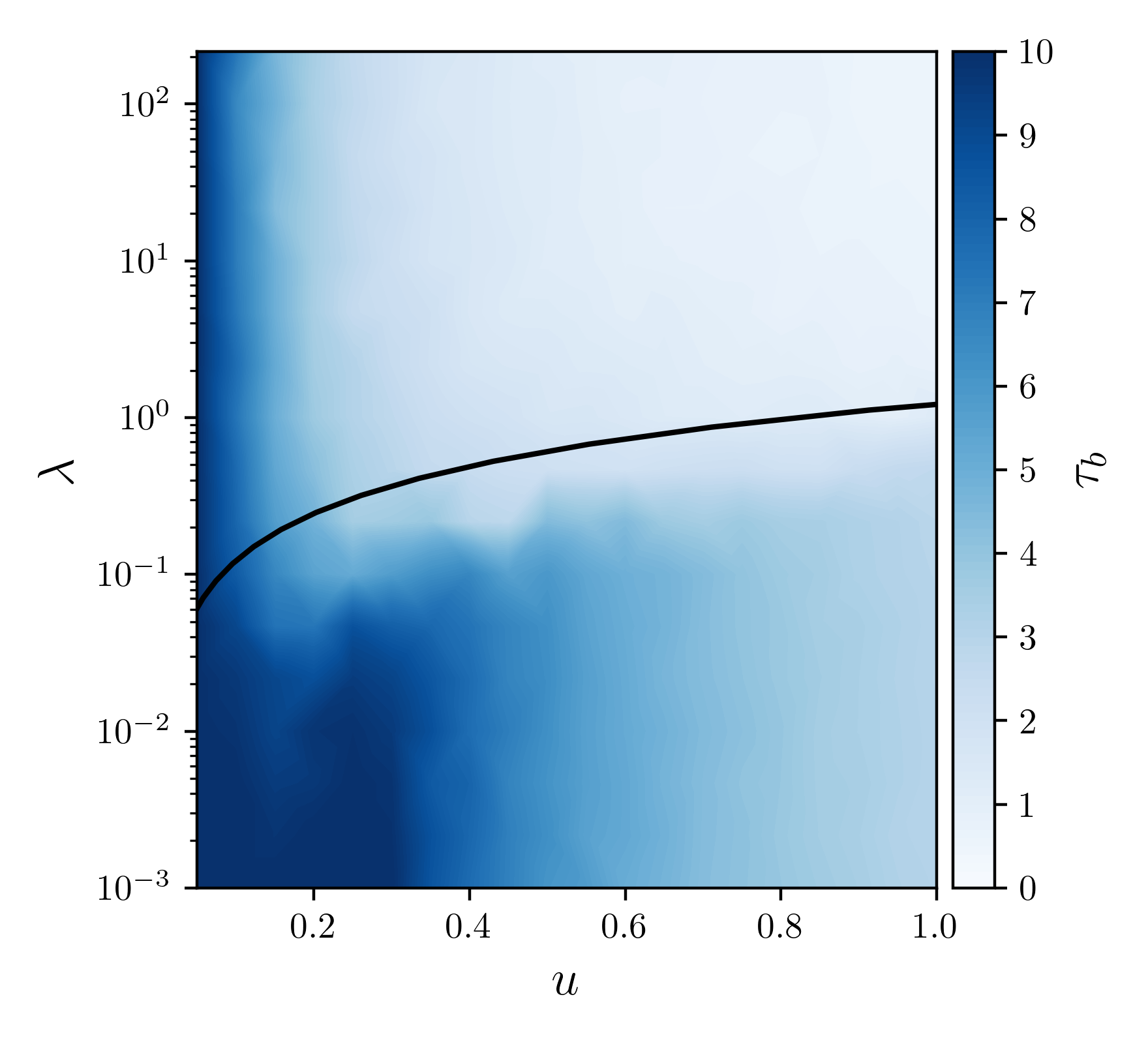}
        \label{fig:stabilization:activation-decay-rate}
    }
    \caption{Including memory in the model is necessary for the description of stable asters. In (a), the geometry of the trailing aster is sketched, highlighting the result obtained in \cref{eq:trailing-distance} in the limit of $k\rightarrow \infty$. In (b), the breakup time \breakuptime is shown for $k=1$ as a function of the activation-pattern velocity \vcirc and the activation decay rate $\lambda$. The solid line represents \cref{eq:lower-critical-velocity} and separates the quasi-stable from the stable region. It can be observed that for long memory (smaller $\lambda$),
    the system can enter a more stable phase increasing the aster's stability since \breakuptime increases. It should be noted that the darkest color indicates $\breakuptime \geq 10$ as no breakup occurred during our simulations.}
    \label{fig:stabilization}%
\end{figure}

It should be noted that \cref{eq:trailing-distance} is only valid if $\frac{\vcirc }{\lambda} - \frac{2r}{\pi}>r$, so that the aster does not reenter the activation area. This condition can be reformulated as a (lower) critical velocity condition
\begin{align}
    \vcirc >\left( 1+\frac{2}{\pi} \right)r\lambda,
    \label{eq:lower-critical-velocity}
\end{align}
differentiating between a stable and a quasi-stable aster in the following sense: If \cref{eq:lower-critical-velocity} is fulfilled, the aster trails at a constant distance behind the activation area, and the aster is indefinitely stable in this simplified treatment. This situation is sketched in \cref{fig:stabilization:sketch}. However, if \cref{eq:lower-critical-velocity} is violated, the aster can reenter, and the calculation leading to \cref{eq:trailing-distance} is no longer valid since $m(t)$ no longer has the simple solution presented in \cref{eq:sol-mass-phase2}. In this situation, we expect that the aster oscillates around the activation area's trailing edge by gaining mass, therefore leaving the activation area, then losing mass, and therefore reentering the activation area. Because the aster will momentarily lose mass (upon leaving the activation area), we describe this regime not as stable but as quasi-stable.

Irrespective of the regime, we have shown that even for small activation retention, the aster persists. Furthermore, given a large enough velocity, the aster trails the activation area. Both these observations are in agreement with the experimental observations~\cite{Ross2019}. Up to now, no differentiation between a quasi-stable and a stable regime has been reported. Direct experimental confirmation based on the activated mass is challenging since the experiments are not sensitive to the activated mass but to light absorbed by the microtubules. However, future experiments might be able to differentiate the two regimes based on the trailing distance, which, in one case, increases linearly with velocity, and in the other case, should oscillate around $r$. 

Additionally, the experiment observed a maximal velocity of the activation area \activationarea, after which no aster is formed, which is not observed in the simple calculation. This difference can be understood in the following way: If the timescale for aster contraction is longer than the illumination time, then the aster cannot form. In our model calculation, however, the contraction is instantaneous due to the assumption of infinite interaction strengths. Therefore, this assumption has to be relaxed. 

Analogous to the previous section, the approximation of infinitely strong interactions can be relaxed using simulations. An example for the simulation of a stable aster is provided in the supplemental material~\footnote{In particular in the file ``stable-aster.gif''}, demonstrating that a stable aster can be formed in our model. 
To further quantify the behavior, we use the breakup time \breakuptime as defined in \cref{eq:def-breakuptime}, to measure the system's stability for systems with finite spring constants. 
We present the simulation results of $\breakuptime (\vcirc ,\lambda)$ for a system with $k=1$ in \cref{fig:stabilization:activation-decay-rate}, where we additionally overlay a solid black line for the condition \cref{eq:lower-critical-velocity}, distinguishing stable from quasi-stable asters. While \cref{eq:lower-critical-velocity} was derived for $k\rightarrow\infty$, the underlying $\breakuptime$ data seems to indicate that this condition is a good approximation even if $k$ becomes finite. Above the solid-black line, only quasi-stable asters exist, and the mass drops immediately upon asters leaving the activation zone. Hence, the breakup time describes the first time the aster leaves the activation area so that $\breakuptime$ is independent of $\lambda$. In contrast, for stable asters trailing the activation area, the breakup time \breakuptime can be considerably longer, but not infinite as for the infinite interaction strength case. Indeed, the now finite aster contraction velocity competes with the activation pattern velocity, which makes asters less stable at higher velocities, which can be observed in our data as well as the experiment.

\section{Description of the Aster-Aster interaction geometry} \label{sec:interaction}
\begin{figure}
    \centering
    \subfloat[][]{
        \includegraphics[width=.8\columnwidth]{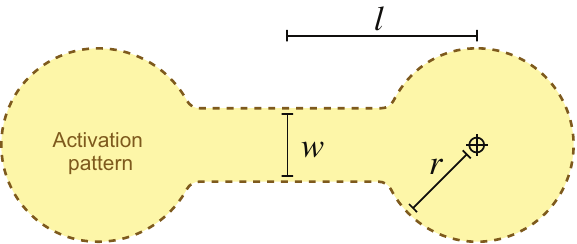}
        \label{fig:dumbbell:geometry}
    }\\
    \subfloat[][]{
        \includegraphics[width=\columnwidth]{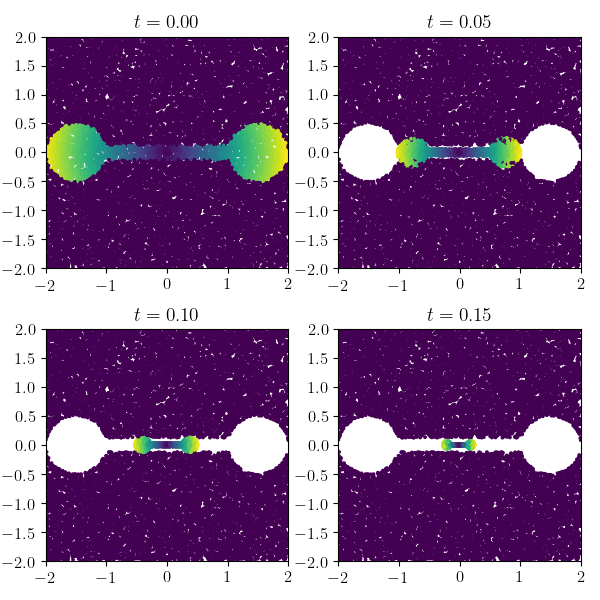}
        \label{fig:dumbbell:snapshots}
    }\\
    \subfloat[][]{
        \includegraphics[width=.9\columnwidth]{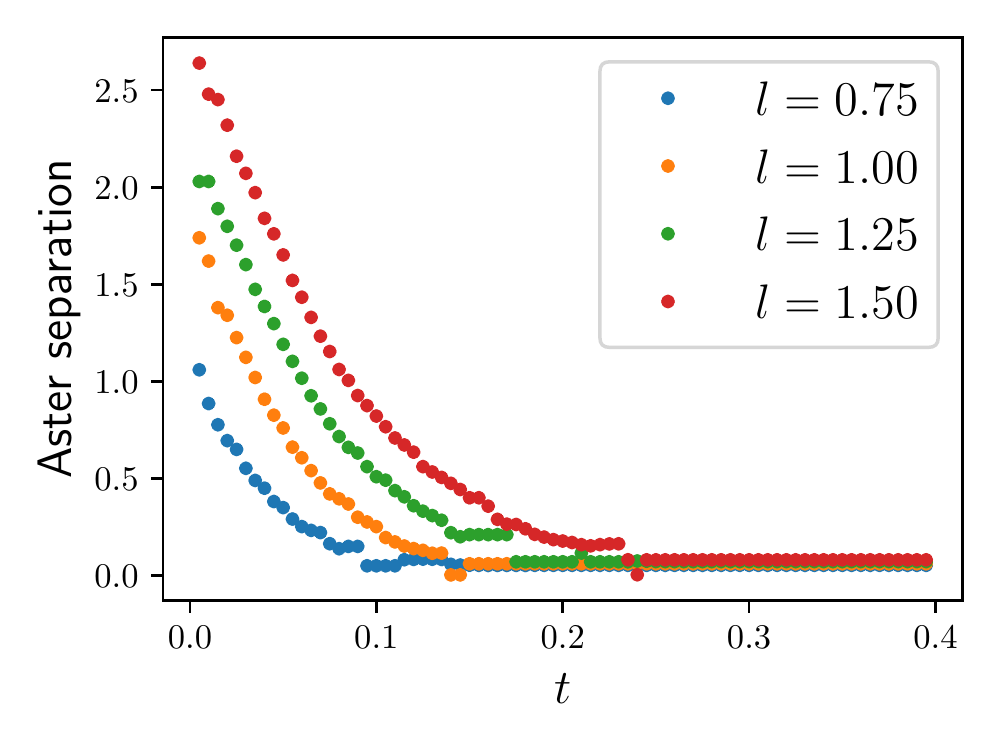}
        \label{fig:dumbbell:distance}
    }
    \caption{Simulation of the model for interacting asters. In (a), the geometry of the simulation is depicted, illustrating the parameters $r$, $w$, and $l$. In (b), illustrations of the system configuration are displayed, illustrating the time evolution of the system for $r=0.5$, $w=0.25$, and $l=1.25$. Here, particles are again colored by their velocity, ranging from violet if they are at rest to yellow for the highest velocity particles. Finally, in (c), the separation between the asters as a function of time is shown for several different separations of activation circles $2l$. As observed experimentally~\cite{Ross2019}, the aster-aster separation seems to decay exponentially fast.}
    \label{fig:dumbbell}%
\end{figure}

As a second example for the applicability of our model, we apply it to simulate the aster-aster interaction. In analogy to the experimental setup probing the aster-aster interactions~\cite{Ross2019}, we consider a geometry as sketched in \cref{fig:dumbbell:geometry}. The purpose of this geometry is to form initial aster is the two large circular areas which then interact via the connecting ``bridge'' that facilitates the merger. 

This interaction geometry is simulated with the model presented in \cref{eq:eom}, for $20\cdot 10^3$ particles, with a time step of $\Delta t=5\cdot 10^{-3}$, and $\lambda=0$ so that once activated particles always remain activated. Otherwise, identical simulation parameters as for the previous simulations are applied. Furthermore, an activation circle radius of $r=0.5$, a bridge width $ w=0.25$, and a half-center distance $l=1.5$ are used. Snapshots of the system (\cref{fig:dumbbell:snapshots}, additional videos with different $l$ in the supplemental material) demonstrate that this activation pattern leads to the merger of two asters as observed in the experiment~\cite{Ross2019}.

To quantify this behavior and to compare it with the experiment, we want to determine the aster-aster distance as a function of time. To measure the pairwise distance between aster cores, we construct the histogram of the $x$-components of all particles at a fixed time $t$. We then fit the histogram with the (heuristic) function
\begin{align}
    h(x) &= A+B\left[ \exp\left( -\frac{(x-x_0)^2}{\sigma^2} \right)+\exp\left( -\frac{(x+x_0)^2}{\sigma^2} \right) \right]\nonumber\\&\phantom{=A}+C \chi_{\left[ -x_0, x_0 \right]}(x),
    \label{eq:distance-fit}
\end{align}
for the parameters $A$, $B$, $C$, $x_0$, and $\sigma$. The parameter $A$ denotes a constant background, and the term proportional to $C$ determines an additional contribution to the linking bridge. In this linking bridge term, $\chi_{[-x_0,x_0]}$ is the characteristic function of the interval $[-x_0,x_0]$, which returns $1$ for arguments in the interval from $-x_0$ to $x_0$ and $0$ otherwise. The term proportional to $B$ denotes the two peaks arising from the asters. Here, we used that the center of the system is aligned with the origin of the simulation. It should be noted that the fit cannot be used in the first time step since the asters have not yet formed. 

The results for the aster-to-aster distance $2x_0$ using this fitting procedure are presented in \cref{fig:dumbbell:distance} for several different activation-function distances $l$ as a function of time. This figure can be compared with Fig.~2e from Ref.~\cite{Ross2019}, where several qualitative agreements can be observed: Both in the experiment and the model, the asters merge, indicated by a separation decreasing to 0. Furthermore, all curves remain ordered by their initial distance, as observed in the experiment. Finally, the decrease of aster separation seems to be exponential in the model and also seems to be exponential in the experiment, at least at later stages. However, the experimental decay seems to be tapered off initially, not observed in the model. This difference might arise due to a finite interaction range in the experiment, in contrast to using the limit $\Cutoff \rightarrow \infty$ in \cref{eq:eom:fp} in the simulations. Nevertheless, a great phenomenological agreement can be observed between the experiment and the simulations, also for the aster-aster interaction geometry.

\section{Conclusions}
In conclusion, in this paper, a novel point-like-particle model was introduced to describe the phenomenology of light-activated microtubule-kinesin systems~\cite{Ross2019}. The model possesses sufficiently elementary limits to derive analytic results while at the same time being versatile enough to include further effects, such as retained activation. Our analysis of the model reveals that the transport arises due to the conservation of the center of mass of activated particles, which can be shifted using the spatiotemporal modulation of interactions, leading to a controlled mechanism for motility. Furthermore, the calculations demonstrate that aster motility is limited due to the mass accumulation without introducing memory. However, if the system is amended with memory in terms of activation retention, asters can persist, possibly even indefinitely. Therefore, including memory in this model is essential to describe experimental systems adequately.

Additionally, simulations were used to verify the model and to analyze applications such as aster motility or aster-aster interactions. For all applications, a good agreement with the phenomenology of the experiments~\cite{Ross2019} could be observed. For example, we demonstrated how the linearly scaling trailing distances arise for motile asters, and we observed the exponential velocity scaling during aster-aster merging. These observations demonstrate our model's applicability to gain insight into this novel class of systems.

Hence, we have shown that the model proposed in this paper can be used to explain the phenomenology of experiments, such as those presented in Ref.~\cite{Ross2019}. Therefore, protein-based active matter and our interaction-based model are an ideal starting point for systematically analyzing microscopic effects emerging from self-organization. Due to the model's versatility, various other features could be analyzed. For example, more realistic particle-particle interactions or the finite extent of particles could be analyzed. Moreover, we limited our discussion to two possible activation patterns \activationarea, inspired by recent experiments~\cite{Ross2019}. However, various other patterns could be imagined~\cite{Stenhammar2016,Qu2020a}, which could lead to exciting light-activated matter systems. 

In such situations, however, our model might require an extension to include fluid dynamics. While we could describe the phenomenology well for the situations described in Ref.~\cite{Ross2019}, recent experiments have studied the interaction between active matter and the surrounding fluid~\cite{Qu2020a}. If these interactions become relevant, the phenomenological model presented here has to be amended with fluid-matter interactions. Nevertheless, we expect our model to serve as an intuitive guide for designing applications and to enable rapid prototyping of novel systems.


Our implementation used several scientific libraries~\cite{Hunter2007,Lam2015,DaskDevelopmentTeam2016,Greff2017}
and is available online~\footnote{Source code available from \url{https://github.com/domischi/SpringBox}}. The data presented in the figures is openly available from the Caltech Research Data Repository at~\footnote{\url{http://dx.doi.org/10.22002/D1.1451}}.
\begin{acknowledgments}
    We thank Jerry Wang, Arjuna Subramanian, Dr.\ Zijie Qu, and Dr.\ Shahriar Shadkhoo for scientific discussions and feedback on the manuscript. 
    We thank Inna-Marie Strazhnik for help preparing some of the figures.
    Furthermore, we acknowledge funding through the Foundational Questions Institute and Fetzer Franklin Fund through FQXi 1816, the Packard Foundation (2019-69662), and the Heritage medical research institute.
\end{acknowledgments}


\bibliography{library-overleaf}

\end{document}